\documentclass[useAMS,usenatbib]{mn2e}

\usepackage{url}
\usepackage[section]{placeins}
\usepackage{graphics,graphicx,times}
\usepackage{pdflscape}
\usepackage{amsmath} 
\usepackage{amssymb}
\usepackage{natbib,astrojournals}

\newcommand{\nii}{[N \textsc{ii}]}
\newcommand{\sii}{[S \textsc{ii}]}
\newcommand{\oi}{[O \textsc{i}]}
\newcommand{\oiii}{[O \textsc{iii}]}
\newcommand{\brg}{Br$\gamma$}

\newcommand{\molhy}{H$_2$}
\newcommand{\ha}{H$\alpha$}
\newcommand{\hb}{H$\beta$}

\def\ltsima{$\; \buildrel < \over \sim \;$}
\def\simlt{\lower.5ex\hbox{\ltsima}}
\def\gtsima{$\; \buildrel > \over \sim \;$}
\def\simgt{\lower.5ex\hbox{\gtsima}}

\title[Shocks in IRAS F17207-0014]{
Shocked Gas in IRAS F17207-0014: ISM Collisions and Outflows}

\author[Medling et al.]
{Anne M. Medling$^1$\thanks{anne.medling@anu.edu.au}, Vivian U$^2$, Jeffrey A. Rich$^3$, Lisa J. Kewley$^1$, Lee Armus$^4$,
\newauthor  Michael A. Dopita$^{1,5,6}$, Claire E. Max$^7$, David Sanders$^6$, Ralph Sutherland$^1$\\
$^1$Research School of Astronomy \& Astrophysics, Mount Stromlo Observatory, Australia National University, Cotter Road, Weston, ACT 2611, Australia \\
$^2$Department of Physics and Astronomy, University of California, Riverside, 900 University Avenue, Riverside, CA 92521, USA \\
$^3$Observatories of the Carnegie Institution of Washington, 813 Santa Barbara Street, Pasadena, CA 91101, USA \\
$^4$Spitzer Science Center, California Institute of Technology, 1200 E. California Blvd., Pasadena, CA 91125, USA\\
$^5$Astronomy Department, King Abdulaziz University, P.O. Box 80203, Jeddah, Saudi Arabia \\
$^6$Institute for Astronomy, University of Hawaii, 2680 Woodlawn Drive, Honolulu, HI 96822, USA \\
$^7$Department of Astronomy \& Astrophysics, University of California, Santa Cruz, CA 95064, USA}

\begin{document}
\date{Accepted 2015 January 5.  Received 2014 December 9; in original form 2014 August 12}
\label{firstpage}
\maketitle

\begin{abstract}
We combine optical and near-infrared AO-assisted integral field observations of the merging ULIRG IRAS F17207-0014 from the Wide-Field Spectrograph (WiFeS) and Keck/OSIRIS. The optical emission line ratios \nii/\ha, \sii/\ha, and \oi/\ha~reveal a mixing sequence of shocks present throughout the galaxy, with the strongest contributions coming from large radii (up to 100\% at $\sim$5 kpc in some directions), suggesting galactic-scale winds. The near-infrared observations, which have approximately 30 times higher spatial resolution, show that two sorts of shocks are present in the vicinity of the merging nuclei: low-level shocks distributed throughout our field-of-view evidenced by an \molhy/\brg~line ratio of $\sim$0.6--4, and strong collimated shocks with a high \molhy/\brg~line ratio of $\sim$4--8, extending south from the two nuclear disks approximately 400 pc ($\sim$0\farcs5). Our data suggest that the diffuse shocks are caused by the collision of the interstellar media associated with the two progenitor galaxies and the strong shocks trace the base of a collimated outflow coming from the nucleus of one of the two disks.

\end{abstract}


\section{Introduction}

Shocks play a significant role in our understanding of galaxy evolution. Widespread shocks in galaxy mergers trace galaxy-scale gas outflows and inflows that can alter the star formation history of a galaxy and their emission can confuse measurements of other physical properties in host galaxies.
Gas can be shocked at the interface between two kinematically distinct gas clouds, wherever supersonic motions occur.  For example, 
cloud-cloud collisions \citep*[e.g.][]{Savedoff67}, tidally-induced gas streams \citep{McDowell03,Colina04,MonrealIbero06,MonrealIbero10,Farage10,Zakamska10}, and
outflows and galactic winds from AGN or star formation \citep[e.g.][]{Heckman90, Lehnert96, VeilleuxRupke02,Lipari04, Sharp10} can all produce radiative shocks to dissipate their kinetic energy \citep{Allen08}.  Indeed, the simulations of \citet{Cox04,Cox06} indicate that a significant fraction of the orbital energy in a galaxy merger will go into shock-heating the gas, potentially driving additional galactic winds.

With such a wide variety of causes for shocks, naturally their effects can be seen in a number of different types of host galaxies, such as galaxy mergers \citep{vanderwerf93}, AGN, and/or starbursts \citep[e.g.][]{Veilleux03}.
Shocks can enhance a galaxy's \nii/\ha, \sii/\ha, and \oi/\ha~emission line ratios by more than 0.5 dex \citep{Allen08}, which can have a number of confounding effects.  Comparisons of these line ratios and \oiii/\hb, used as diagnostics of the energy source \citep*[as in the BPT/VO87 diagrams;][]{BPT, Veilleux87}, would misidentify a starburst galaxy with shocks as a LINER or composite galaxy \citep[e.g.][]{Armus89,Veilleux95,Martin97,Dopita97,Kim98,Allen99,Veilleux99,Sharp10,Rich10,Rich11,Rich14}. Measurements that rely on emission line fluxes or ratios, such as star formation rate and metallicity, can also be affected unless shocks are identified and removed from the calculation.

Shock-excited molecular gas can also be observed at near-infrared wavelengths, with a series of ro-vibrational emission lines of \molhy~in the range $\sim$1.9--2.4 $\mu$m \citep{Mouri94}. These have been observed both in nearby supernova remnants \citep*[e.g.][]{Oliva89} and in extragalactic shocks \citep{Veilleux97,Sugai97,Sugai99,Sugai03,Davies00,Veilleux09,Zakamska10,Hill14,Davies14_H2}. Additional ro-vibrational transitions in the mid-infrared have also been used to trace such shocks \citep{Appleton06,Peterson12,Mazzarella12,Cluver13}.

Locally, Luminous and Ultra-Luminous InfraRed Galaxies (U/LIRGs, with log($L_{IR}/L_{\sun}) \geq 11$ for LIRGs and $\geq 12$ for ULIRGs) are excellent laboratories to study shocks, as $L_{IR}$ correlates with star formation rate, AGN fraction, and merger fraction \citep{Sanders88,Veilleux95,Sanders96,Veilleux02,IshidaPhD}, each of which can produce shocks. Outflows are commonly observed in local U/LIRGs \citep{Rupke02,RupkeVeilleux05,Rupke05c,Rupke05a,Rupke05b,Martin06,Sturm11,RupkeVeilleux13,Veilleux13,Teng13,Spoon13,RodriguezZaurin13,Arribas14,Cicone14}, and are likely a common cause of shocks. Several studies of shocks in nearby galaxies have linked them to superwinds produced by stellar feedback \citep{Heckman90,Grimes05,Rich11,Soto12a,Soto12b,Hopkins12,Hong13} and AGN-driven outflows \citep{Mazzarella12}. Although these systems are relatively rare in the local universe, \cite{LeFloch05} have shown that high infrared luminosities may have been the norm at redshifts beyond 0.7. To understand these cosmically important objects and the physical processes that occur in them, the Great Observatory All-Sky LIRG Survey \citep[GOALS;][]{Armus09} team has compiled data across a number of different wavelength regimes, including Hubble Space Telescope imaging in the optical \citep{Kim13} and near-infrared \citep{Haan11}, Chandra X-ray observations, \citep{Iwasawa11}, Spitzer spectra in the mid-infrared \citep{Stierwalt13}, GALEX observations in the near and far ultraviolet \citep{Howell10}, and Herschel Space Telescope spectra in the far-infrared \citep{DiazSantos13}.

Recently, subsamples of the GOALS galaxies have been studied with integral field spectroscopy, allowing an unprecedented look at the physical mechanisms present in U/LIRGs. In \citet{Rich14}, a sample of 27 U/LIRGs was observed with the the Wide Field Spectrograph WiFeS to obtain resolved optical spectroscopy (R$\sim$3000 and 7000 in the blue and red sides, respectively). The high spectral resolution afforded by the WiFeS instrument enabled the decomposition of emission line ratios and revealed the prevalence of shocks, particularly in late merger stages, \citep{Rich10,Rich11,Rich14} which cause star-forming galaxies to appear as LINERs or composite galaxies in a single-aperture analysis. Though IFU data are capable of identifying the presence of shocks and their relative contribution to the total emission line flux, WiFeS lacks the spatial resolution and the long wavelengths necessary to trace shocks in the dusty cores of U/LIRGs.

We present the first results of our new study that combines two sets of integral field spectroscopy: optical observations with a wide field-of-view and moderately high spectral resolution to identify the presence of the shocks, and near-infrared high spatial resolution observations to investigate the origin of shocks in the cores of nearby U/LIRGs. This paper focuses on IRAS F17207-0014, the first target for which we have obtained both datasets.

IRAS F17207-0014 (log($L_{IR}/L_{\sun}$) = 12.46, 17h23m22.01s -00d17m00.2s, z=0.043, 0.848 kpc arcsec$^{-1}$) is a late-stage major merger \citep[merger stage classification 5 out of 6, `single or obscured nucleus with long tidal tails';][]{Haan11, Kim13}, and the most infrared luminous system in the southern sky. Although dust obscuration in the optical and limited spatial resolution in the near-infrared ($\frac{\lambda}{D}\sim$0\farcs15) have prevented Hubble observations from separating the two nuclei, new adaptive optics observations with the OH-Suppressing InfraRed Imaging Spectrograph (OSIRIS) have shown that the two nuclei are still distinct \citep{nucleardisks}, though they overlap with a small central separation in projection (0\farcs24 $\sim200$ parsecs). 
We note that \citet{Martin06}, \citet{Soto12a}, and \citet{Soto12b} classified this galaxy as a double nucleus separated by 2--3\arcsec based on kinematics measured from their longslit data.  Our OSIRIS data reveal that the two nuclei are considerably closer, and it is thus possible that the second nucleus seen by previous authors was in fact merely a second kinematic component (e.g. tidal stream).
This galaxy is classified as a star forming system based on observations in the optical \citep{Veilleux95,Yuan10}, mid-infrared \citep{Lutz99,Risaliti06,Stierwalt13,Inami13}, and X-ray \citep{Franceschini03,Iwasawa11}.  

This galaxy has been studied by several other authors to date, most finding evidence for outflows on a variety of scales. \citet{Martin06}, using long-slit optical spectroscopy, found Na I D absorption blueshifted by $\sim$400 km s$^{-1}$ relative to the \ha~emission. In a similar analysis, \citet{Rupke05a,Rupke05b} found Na I D absorption blueshifted from systemic by $\sim$300 km s$^{-1}$, also interpreting this as outflowing neutral gas. \citet{Sturm11} looked at Herschel/PACS spectra of the OH 79 $\mu$m line and found evidence of blueshifted/outflowing molecular gas with a velocity offset of -100 km s$^{-1}$ (v$_{85}$ = -170 km s$^{-1}$, v$_{max}$ = -370 km s$^{-1}$). A number of authors have examined optical emission lines for signs of outflows as well.  \citet{Arribas03} found no evidence of outflows in their INTEGRAL observations (although they do suggest an inflowing component to the southwest); \citet{Westmoquette12} also find no evidence for outflows in their VIMOS data.  However, \citet{Soto12a} fit multiple components to the emission lines in their optical long-slit spectra and produce measurements of velocity offset ($\sim$-100 km s$^{-1}$ from systemic) and velocity dispersion ($\sim$100 km s$^{-1}$) for the shocked components, which they sum together to produce an estimated outflow velocity of 200 km s$^{-1}$; they also use emission line ratios and shock models to infer shock velocities ranging from 50--600 km s$^{-1}$.  \citet{Arribas14} looked again at INTEGRAL data for this object and found, integrated over the galaxy, an outflowing component to the emission lines, with v$_{offset} = -151\pm82$ km s$^{-1}$ and v$_{max} = -306\pm87$ km s$^{-1}$.

In this paper, we present an updated interpretation of shocks in IRAS F17207-0014 from the combination of wide-field optical and AO-assisted near-infrared integral field spectroscopy.
In Section~\ref{obs} we describe the observations and line-fitting techniques.  
In Section~\ref{models} we include the details of our shock models used for optical emission line analysis.
In Section~\ref{results} we present the analysis from both our optical and near-infrared data. In Section~\ref{discussion} we discuss the implications of the combined data and in Section~\ref{summary} we summarize the results. Throughout the paper we adopt $H_0 = 70$\,km\,s$^{-1}$\,Mpc$^{-1}$,
$\Omega_{\rm m}$ = 0.28, and $\Omega_\Lambda$ = 0.72 \citep{Hinshaw09}.  Physical scales were calculated using Ned Wright's Cosmology Calculator\footnote{Available at \url{http://www.astro.ucla.edu/~wright/CosmoCalc.html}.} \citep{Wright06}.


\section{Observations and Line Fitting }
\label{obs}

As a pilot study for our larger survey, we compare here two sets of integral-field observations of IRAS F17207-0014. Our optical integral field spectroscopy was taken with the Wide Field Spectrograph (WiFeS) as part of a larger survey of shocks in galaxy mergers. A more detailed analysis of its shocks and kinematics, along with a comparison to the population of mergers, will be presented in Rich et al. (in prep). 
The near-infrared integral field spectroscopy was first presented in \citet{nucleardisks} as part of a survey studying nuclear kinematics in gas-rich mergers.

\subsection{WiFeS Data}
\begin{figure}
\centering
\includegraphics[scale=0.5]{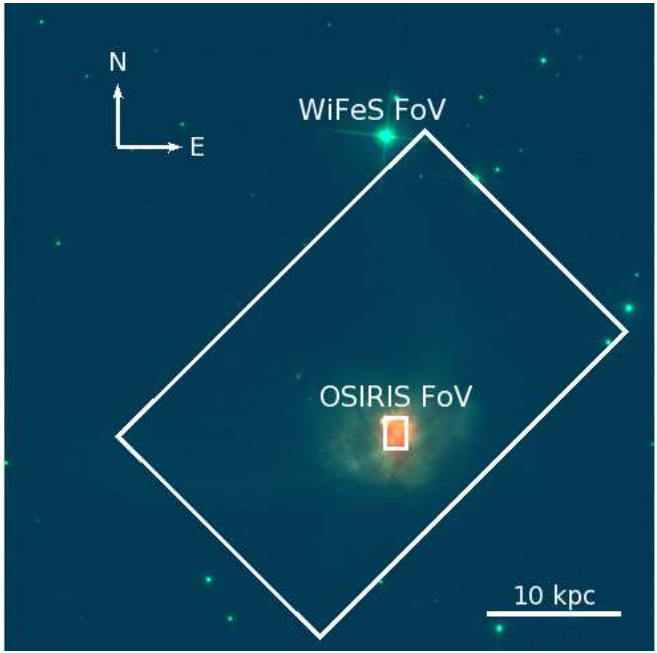}
\caption{Three-colour image of IRAS F17207-0014 from Hubble ACS $F435W$ (blue), ACS $F814W$ (green), and NICMOS $F160W$ (red) from \citet{Kim13} and \citet{Haan11} showing the field of view for the WiFeS data (large white box, 25\arcsec $\times$ 38\arcsec) and the OSIRIS $Kcb$ data (small white box, 1\farcs8 $\times$ 2\farcs7).  Images are shown with a square-root flux scale in order to emphasize outer structure.}  
\label{FOV}
\end{figure}

The first set of data was taken with the Wide Field Spectrograph \citep[WiFeS;][]{Dopita07,Dopita10} on the 2.3-meter telescope at Siding Spring Observatory. WiFeS is an image slicer with dual beams, enabling red and blue spectra to be obtained simultaneously, with different spectral resolutions and wavelength coverage. The angular resolution is seeing-limited and is $\sim$1.5\arcsec~for our observations, corresponding to $\sim$1.3 kpc for IRAS F17207-0014. The observations were taken on 3 June 2010 UT with the B3000 and R7000 gratings, giving a spectral resolution of R$\sim$3000 in the range 3700--5700\AA~and a resolution of R$\sim$7000 in the range 5700--7000\AA. These data will be presented in full as part of the WIGS survey (Rich et al. in prep).
The data were sampled with square spatial pixels of 0\farcs5 on a side which were then binned $2\times2$ to produce 1\arcsec~spaxels. The total field of view is 25\arcsec $\times$ 38\arcsec, shown in Figure~\ref{FOV}. The data were taken in 3 exposures of 1500 seconds each, yielding a total on-source time of 75 minutes. The data were reduced using the WiFeS data reduction pipeline \citep{Dopita10}. For more details on the observations and reduction process, see \cite{Rich11}.

We obtain emission line fits for this study using \textsc{LZIFU} (Ho et al. in prep), which uses the penalized pixel fitting routine \textsc{pPXF} \citep{ppxf} to fit stellar template models \citep[here we used the library of templates at solar metallicity from][]{gonzalezdelgado05} to the continuum and then fits one, two, or three Gaussian profiles to each emission line. \textsc{LZIFU} fits both the red and the blue spectra simultaneously, allowing for consistent continuum subtraction across the whole spectrum. By fitting both sides simultaneously, \textsc{LZIFU} is also able to fit the same velocity and velocity dispersion for all emission lines, varying only the flux. This fitting method has the beneficial effect of using the high spectral resolution of the red side to inform the line profiles of the blue side, allowing for high fidelity flux ratios for multiple components. Due to signal-to-noise ratio limitations, this dataset was fit with the one-component mode only.

\subsection{OSIRIS Data}
\label{OSIRISobs}
The near-infrared set of data was taken on 23 and 24 May 2011 UT with the OH-Suppressing InfraRed Integral-field Spectrograph \citep[OSIRIS;][]{Larkin06} on the Keck II telescope using laser guide star adaptive optics (LGS AO). OSIRIS uses a lenslet array to obtain up to 3000 spectra at once, at a spectral resolution of approximately R$\sim$3000. The observations of IRAS F17207-0014 were taken in two modes: a) using the $Kcb$ filter (1.965--2.381 $\mu$m) at the 0\farcs1 spaxel$^{-1}$ plate scale, and b) using the $Hn4$ filter (1.652--1.737 $\mu$m) at the 0\farcs035 spaxel$^{-1}$ plate scale. The field of view of the $Kcb$ filter is shown in Figure~\ref{FOV} for comparison; the field of view of the $Hn4$ data is about half that size (see Section~\ref{osiris} for relative position). Data were taken in ten-minute exposures set in an object-sky-object pattern, for total on-source exposure times of 60 and 40 minutes for the $Kcb$ and $Hn4$ filters, respectively.  

OSIRIS resolves below the seeing limit by utilizing Keck's laser guide star adaptive optics system \citep{Wiz00,vanDam04,Wiz06,vanDam06}. This LGS AO system uses a pulsed laser tuned to the Sodium D$_{2}$ transition at 589 nm to excite atoms in the sodium layer of the atmosphere, around 95 km. The distortions of this laser spot are then measured and a deformable mirror is used to flatten the wavefront, correcting for atmospheric turbulence in both the laser spot and the science data in real time. An on-sky source must be available to correct for image motion; this tip-tilt star must be brighter than $\sim$18th magnitude in $R$ and within 1' of the science target for the Keck AO system. For these observations, we used a nearby star of $R$-band magnitude 14.1 and separation of 25\farcs6. The Keck AO system enables diffraction-limited imaging in the near-infrared ($\sim$0\farcs065 in $K$-band).

These OSIRIS data were reduced using the OSIRIS pipeline v2.3\footnote{Available at \url{http://irlab.astro.ucla.edu/osiris/pipeline.html}.}, using the updated OSIRIS wavelength solution for data taken after October 2009 which is now available in version 3 of the pipeline. For more details on the reduction process, see \citet{nucleardisks}.  We flux-calibrated our $Kcb$ datacube using the NICMOS $K$-band image presented in \citet{Scoville00}.  Unfortunately, no equivalent calibration data were available to flux-calibrate our $Hn4$ datacube.

Emission lines in the OSIRIS datacubes were fit as in \citet{mrk273} and \citet{nucleardisks}, with single-Gaussian components. Rather than fit all lines simultaneously, lines were fit according to species, allowing for different kinematic structure in ionized vs. molecular gas tracers. Here we present morphological results from the \molhy~emission, which has a series of 5 emission lines in the $K$-band. For more details on the kinematics and other tracers, see \citet{nucleardisks} and U et al. (in prep).

\section{Shock Models}
\label{models}

In the following sections, we compare our optical emission line ratios to shock models, which we describe here.

The grid of slow shock models used here was computed by \citet{Rich10} from the \textsc{MAPPINGS III} code \citep[updated from][]{Sutherland93}. Briefly, these models are analogous to the fast-shock models presented in \citet{Allen08}. Following \citet{Rich11}, we restrict our models to fully-preionized shocks with shock velocities ranging from 100 to 200 km s$^{-1}$, which match our data most consistently.

The abundance set used in these models is that of \citet{Grevesse10} with standard solar dust depletion factors from \citet*{Kimura03}. We note that using these dust depletion factors is valid only for shocks slow enough not to sputter the dust grains which may be advected into the shocked region. The models were then calculated for a range of metallicities, ranging from log(O/H)+12 = 7.39 to 9.39; only models with log(O/H)+12 = 8.69 or higher were considered for IRAS F17207-0014. These models also include a transverse magnetic field consistent with equipartition of the thermal and magnetic fields ($B \propto n_{c}^{-1/2}$; $B = 5 \mu$G for $n_{c} = 10$ cm$^{-3}$).  We note that this choice of magnetic field is consistent with the findings of \citet{Thompson06}, who found that starbursts likely have magnetic fields considerably higher than the minimum magnetic field required to produce radio emission.  The flux ratio of the doublet \sii6716/\sii6731 is sensitive to magnetic field strength, and thus can be used as a check of assumptions.  We find that our equipartition choice of model predicts a ratio \sii6716/\sii6731 of 1.0-1.2, whereas shock models using a much smaller magnetic field predict a ratio of 0.5-0.75.  Our observations have \sii6716/\sii6731 $\sim$1.3, more consistent with a magnetic field in equipartition.

For shocks of each metallicity and velocity, the emission line ratios \oiii/\hb, \nii/\ha, \sii/\ha, and \oi/\ha~were calculated for comparison with the observed emission line ratios (see Section~\ref{wifes}).


\section{Results: Line Ratios and Maps}
\label{results}

\subsection{WiFeS Analysis}
\label{wifes}

The emission line ratios \nii/\ha, \oiii/\hb, \sii/\ha, and \oi/\ha~are used as diagnostic tests of the ionization mechanisms present in the gas. Different physical processes, e.g. HII regions vs. AGN, will cause different flux ratios of these \textbf{transitions}, and therefore objects fall in different locations in the standard diagnostic diagrams depending upon their primary source of excitation.
The hard radiation field from an AGN produces more excitations of [OIII], [NII], [SII] and [OI] compared with the thermal radiation field from the young hot O and B stars in HII regions \citep{BPT, Veilleux87, Kewley01, Kauffmann03, Kewley06}.
Though the diagnostic diagrams are commonly used to distinguish between star-forming galaxies and AGN hosts, shocks can elevate emission line ratios, causing a spectrum to move from the HII region of these line diagnostics towards the composite or LINER regions \citep{Armus89,Veilleux95,Martin97,Dopita97,Kim98,Allen99,Veilleux99,Soto12b,Hong13,Rich14,Newman14}.  Mixing between AGN and star formation can also exist \citep{Davies14}, but such mixing increases the \oiii/\hb~ratio more strongly, shifting points towards the Seyfert region of the diagnostic diagrams. By comparing these ionization line ratios in spatially-resolved spectroscopy to photoionization and shock models, it is thus possible to map out the shocked regions and identify the relative contribution to the luminosity from shocks, star formation, and AGN \citep{MonrealIbero06,MonrealIbero10,Sharp10,Rich10,Rich11,Westmoquette11,Freeland11,Soto12a,Davis12,Fogarty12,Yuan12,Vogt13,Ho14}.

\begin{figure*}
\centering
\includegraphics[scale=1]{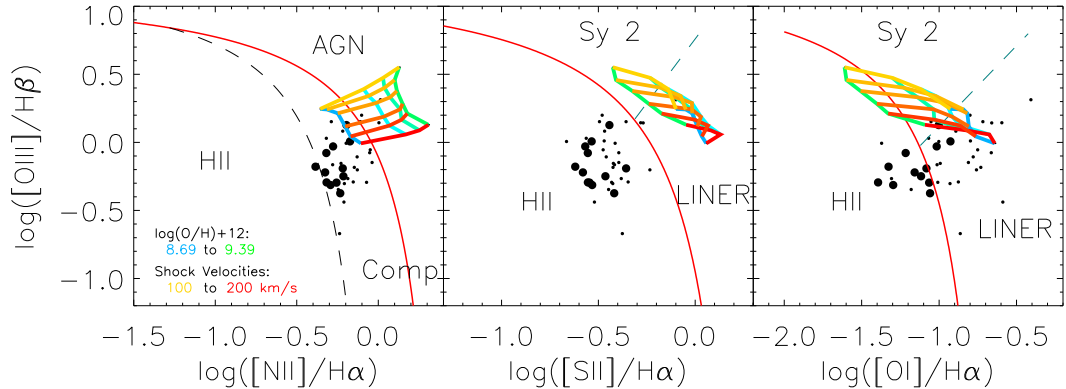}
\caption{Emission line ratio diagnostic diagrams for IRAS F17207-0014. Likely ionization mechanisms for each region of the plot are suggested based on the extreme starburst line \citep[solid red line,][]{Kewley01}, the pure star formation line \citep[dashed black line,][]{Kauffmann03}, and the Seyfert-LINER line \citep[teal dashed line,][]{Kewley06}, though these classifications do not take into account the effects of shocks. Line ratios from single-component fits to each spaxel with highest signal-to-noise ($>$8) are plotted as large black dots and those of signal-to-noise 3--8 as small dots. Pure shock models from an updated version of \textsc{MAPPINGS III} \citep{Sutherland93,Farage10,Rich10} are overlaid, demonstrating that shocks likely contribute to the line ratios of $>$80\% of these spaxels, causing them to fall above the photoionization line in one or more diagnostic plots. The grid includes shock-induced line ratios from shocks with velocities of 100 (yellow, top horizontal grid line), 120, 140, 160, 180, and 200 km s$^{-1}$ (red, bottom horizontal grid line) and for gas with metallicities log(O/H)+12 of 8.69 (blue, vertical grid line on left in first panel and right in others), 8.69, 8.99, 9.17, and 9.39 (green, opposite vertical grid line).  }  
\label{BPTdiagramgrid}
\end{figure*}

In Figure~\ref{BPTdiagramgrid}, we show the standard emission line diagnostic diagrams for IRAS F17207-0014 for spaxels where emission lines in all relevant species have a signal-to-noise ratio of 3 or higher (small black points) and 8 or higher (large black points). We include two signal-to-noise ratio cuts to avoid sensitivity to uncertainties in noise levels in our spectra. All spaxels fall above the pure starburst classification line in the \nii/\ha~diagnostic, indicating that HII regions alone cannot produce the observed line ratios. We note that our line ratios are similar to those measured in the corresponding apertures of the longslit data of \citet[][their Figure 2.28]{Soto12a}. We have overlaid the shock models (coloured grid lines) described in Section~\ref{models}. Our points form a mixing sequence of starburst-dominated (toward the bottom left of the plots and below the Kauffmann pure star formation line) and shock-dominated (towards the right of the plots) spaxels.
These mixing sequences indicate a varying fraction of the emission due to starbursts vs. shocks \citep{Rich10,Rich11}.  

\begin{figure*}
\centering
\includegraphics[scale=1]{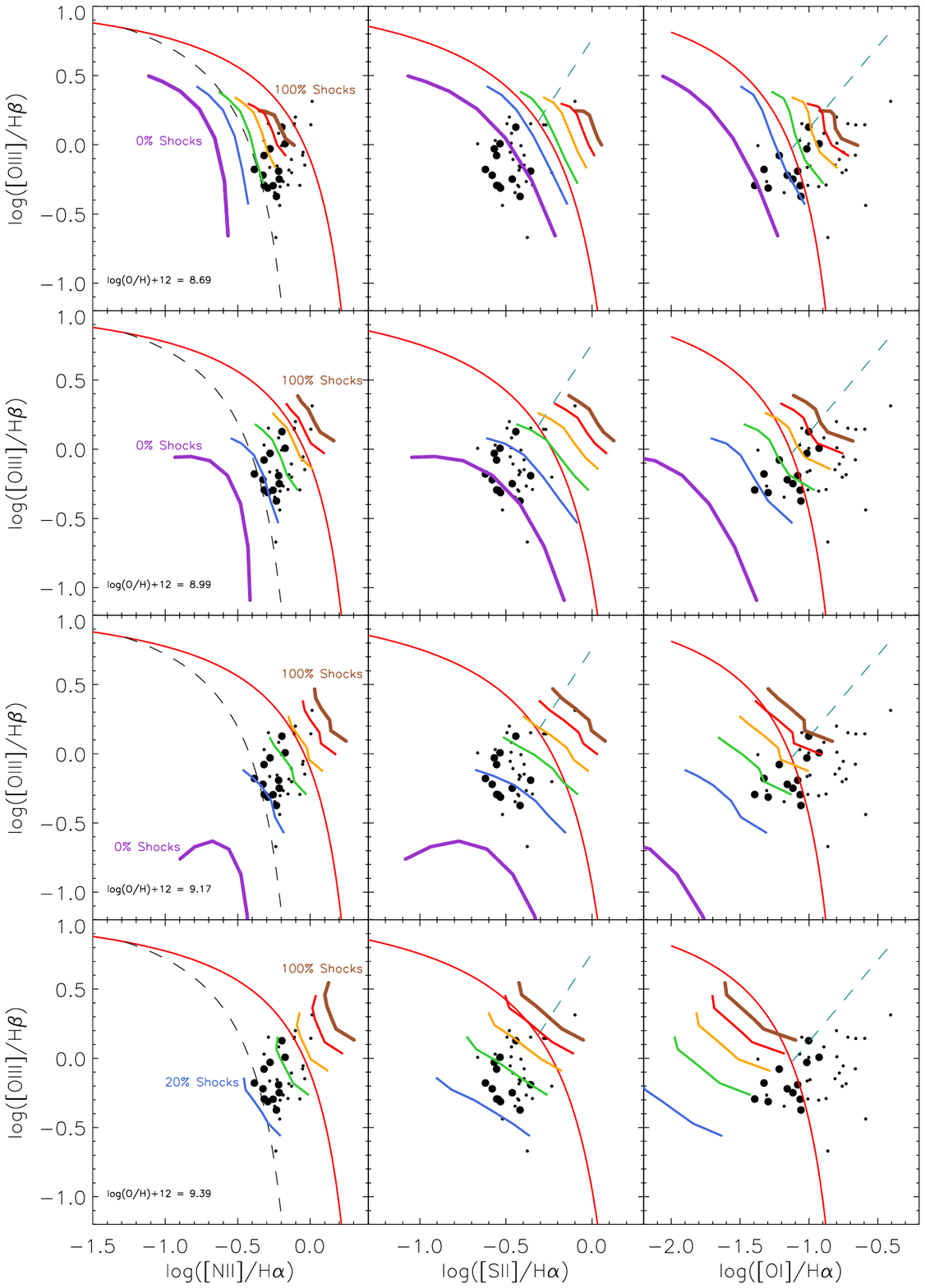}
\caption{Emission line ratio diagnostic diagrams for IRAS F17207-0014. The extreme starburst line \citep[solid red line,][]{Kewley01}, the pure star formation line \citep[dashed black line,][]{Kauffmann03}, and the Seyfert-LINER line \citep[teal dashed line,][]{Kewley06} are the same as in Figure~\ref{BPTdiagramgrid}. Line ratios from single-component fits to each spaxel with highest signal-to-noise ($>$8) are plotted as large black dots and those of signal-to-noise 3--8 as small dots. In each plot, we show line segments indicating models of 0\% shocks (violet) to 100\% shocks (brown) in 20\% increments, with metallicity log(O/H)+12 = 8.69 (top row), 8.99, 9.17, or 9.39 (bottom row). Models with log(O/H)+12 = 8.99 and 9.17 account for all high and most moderate signal-to-noise spaxels.  Between these two models, the lowest line ratios can be produced by $<$20\% shocks (second row) or 20--40\% shocks (third row).}  
\label{mixing}
\end{figure*}

By combining photoionization models from Starburst99 \citep{Leitherer99} and \textsc{MAPPINGS III} with the shock models discussed previously, we plot four possible mixing sequences in Figure~\ref{mixing}, varying the metallicities of the models in each case. That is, for each metallicity, we plot the range of ionization line ratios possible from photoionization models with a range of ionization parameters (log(Q(H)) = 6.5--4e8) combined with shocks of a range of velocities (100--200 km s$^{-1}$).  The lowest and highest metallicity models (log(O/H)+12 = 8.69 in the top row and 9.39 in the bottom row, respectively) do not produce mixing sequences able to account for all high signal-to-noise ($>$8) points, and are ruled out. We note that no models are able to reproduce all points with moderate signal-to-noise ratios ($>$3), particularly those in the log(\oi/\ha) plot.  This could be due to increased noise in the \oi~measurements or to difficulties in the shock models themselves.  We are in the process of producing new shock models (MAPPINGS IV; Sutherland et al. in prep) with updated atomic data that may improve \oi/\ha~ratio predictions.  In the meantime, we caution the reader against using solely \oi/\ha~measurements as shock diagnostics.

As shown in the second row of panels, our mixing models with log(O/H)+12 = 8.99 indicate that the least-shocked spaxels contain 10\% $\pm$ 10\% shocks, because the lowest spaxels fall between the purple and blue lines on each of the diagnostic diagrams. The third scenario, with log(O/H)+12 = 9.17, has the lowest spaxels never reaching the purple line, and sometimes above even the blue line: this suggests that 30\% $\pm$ 10\% of the ionization in these regions comes from shocks. It is encouraging that these mixing sequences are not highly sensitive to the gas-phase metallicity in this regime. 

\begin{figure*}
\centering
\begin{minipage}{4.8in}
\includegraphics[trim = .5cm 0cm 0cm 0cm, scale=0.8]{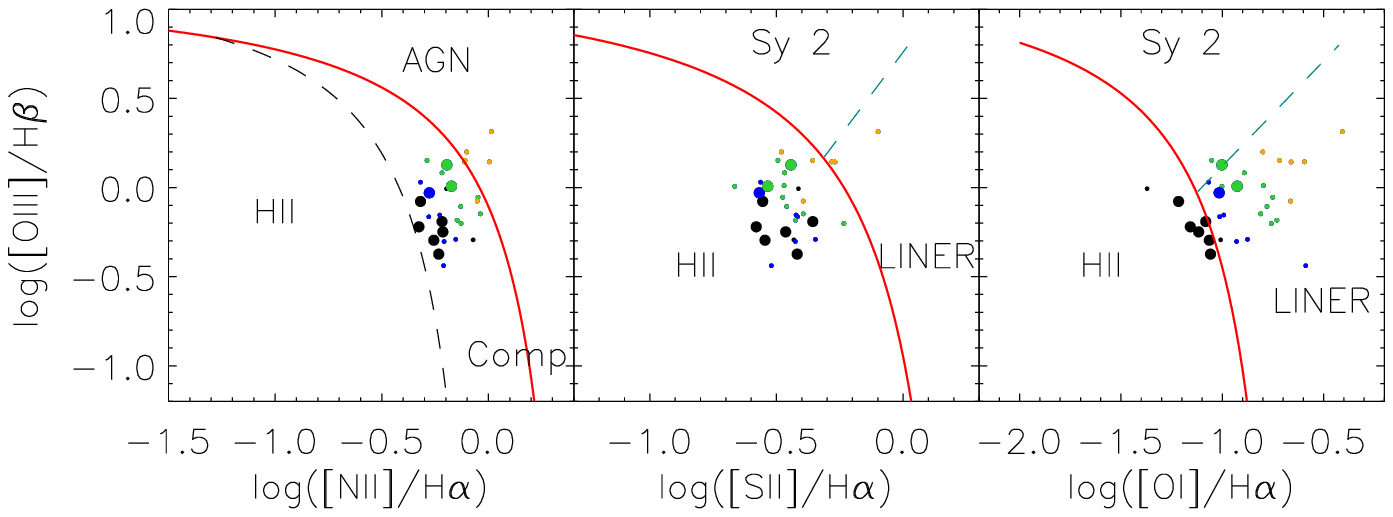}
\end{minipage}
\begin{minipage}{1.6in}
\vspace{1cm}
\includegraphics[trim = 4cm 0cm 0cm 0cm, scale=0.8, angle=-45]{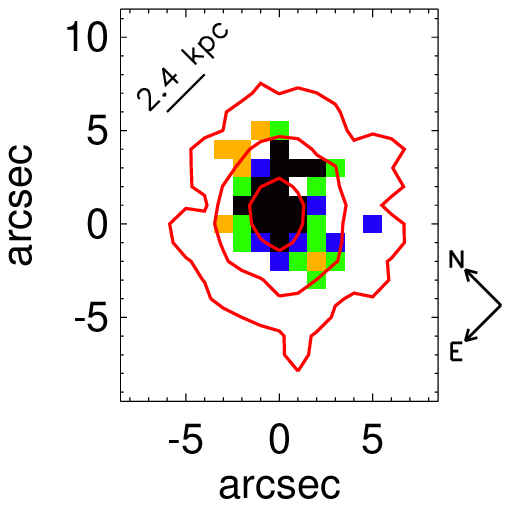}
\end{minipage}
\caption{Emission line ratio diagnostic diagrams for IRAS F17207-0014. The extreme starburst line \citep[solid red line,][]{Kewley01}, the pure star formation line \citep[dashed black line,][]{Kauffmann03}, and the Seyfert-LINER line \citep[teal dashed line,][]{Kewley06} are the same as in Figure~\ref{BPTdiagramgrid}. Line ratios from single-component fits to each spaxel with high signal-to-noise ($>$3 for small points, $>$8 for large points) are plotted in the left three panels, colour-coded according to position in the third plot, with colour divisions evenly spaced in log(\oiii/\hb) + log(\oi/\ha). On the right, we include a zoomed-in image of the galaxy with spaxels colour-coded in the same fashion, and with red contours in \ha~flux overlaid to give scale. White pixels indicate regions with signal-to-noise $<3$ for one or more key diagnostic emission lines.  We see that the central spaxels have the strongest contribution from star formation (60--100\%), with outer spaxels having a higher relative contribution from shocks (80--100\%).}  
\label{reverseBPT}
\end{figure*}

We also look at the spatial locations of the spaxels with the lowest contribution from shocks in Figure~\ref{reverseBPT}. Here we have colour-coded the spaxels in the BPT diagrams and then included an image of the galaxy to the left, with each spaxel printed in its matching colour and with \ha~flux contours to show perspective.  (\ha~emission is morphologically similar to the continuum emission, with the brightest point indicating the galaxy core.) The colour-coding corresponds to shock fraction, with orange spaxels indicating highest shock fraction. These diagnostics show that the central regions of the galaxy have the highest relative contribution of star formation (60--100\%), and suggest that shocks are dominant at radii beyond a few kiloparsecs, where they contribute 80--100\% of the line emission. Though the relative contribution of shocks (0--40\%) is lower in the centre, this could be due to an increase in star formation rather than a decrease in shocks. To test the idea that shocks are present in the nucleus, we look to higher spatial resolution data in the central regions.

\subsection{OSIRIS Analysis}
\label{osiris}

To investigate the central spaxels of the WiFeS data in more detail, we turn to the OSIRIS data, with a field-of-view that covers only approximately the central 2x2 WiFeS spaxels. The GALFIT modeling and kinematics presented in \citet{nucleardisks} reveal that the core of IRAS F17207-0014 contains the nuclei of the two progenitor galaxies separated by $\sim200$ parsecs in projection. Both nuclei appear to host small nuclear disks ($R_{eff}$ of 200 and 410 parsecs, respectively).  

Though the $K$-band observations are at a lower spatial sampling (0\farcs1 spaxel$^{-1}$) than the $H$-band observations, they are particularly interesting because of the \molhy~emission present, which traces warm molecular gas. 
There are five \molhy~transitions falling within our observed wavelength coverage: 1--0 S(3) at 1.7586 $\mu$m, 1--0 S(2) at 2.0338 $\mu$m, 1--0 S(1) at 2.1218 $\mu$m, 1--0 S(0) at 2.2235 $\mu$m, and 2--1 S(1) at 2.2477 $\mu$m.  (All five of these transitions are seen in our integrated spectrum, although no individual spaxel detects all five transitions.  Throughout the rest of the discussion, when we refer to \molhy~flux, we are referring to the strongest transition, 1--0 S(1) at 2.1218 $\mu$m, unless otherwise specified.)
These vibrational transitions of \molhy~can be excited by shock heating, UV fluorescence or thermal excitation from O and B stars, and/or X-ray heating from a nearby AGN \citep{Mouri94}. Because we do not see a broad \brg~line, and because no AGN is evident from optical \citep{Yuan10}, mid-infrared \citep{Stierwalt13}, or X-ray \citep{Iwasawa11} observations, the first two scenarios are more likely for this galaxy. If \molhy~gas is excited by UV light from massive stars, the light from those stars should also produce \brg~emission. Therefore a high \molhy/\brg~ratio in this case would indicate shocks rather than star formation \citep[see e.g.][]{Goldader97, Tecza00}.  Star-forming regions have \molhy/\brg~ratios $\lesssim$0.6 \citep{Joseph84, Moorwood88,RodriguezArdila04,RodriguezArdila05,Riffel10}; shock-excited Galactic and extragalactic sources have \molhy/\brg~ratios greater (or much greater) than unity \citep{Puxley90}. The scenario of shock-excited \molhy~emission is also supported by key ratios of \molhy~transitions: 1--0 S(3) / 1--0 S(1) = 0.92, 1--0 S(2) / 1--0 S(0) = 1.6, and 2--1 S(1) / 1--0 S(1) = 0.12 all fall within the regime predicted by the shock models of \citet{ShullHollenbach78} and \citet{Chernoff82} and are similar to those predicted by \citet{Mouri94} using the model of \citet{Brand89}. We see no difference in our measured ratios between the high \molhy/\brg~ratio region and the rest of the nuclear regions.

\begin{figure}
\centering
\includegraphics[scale=.6]{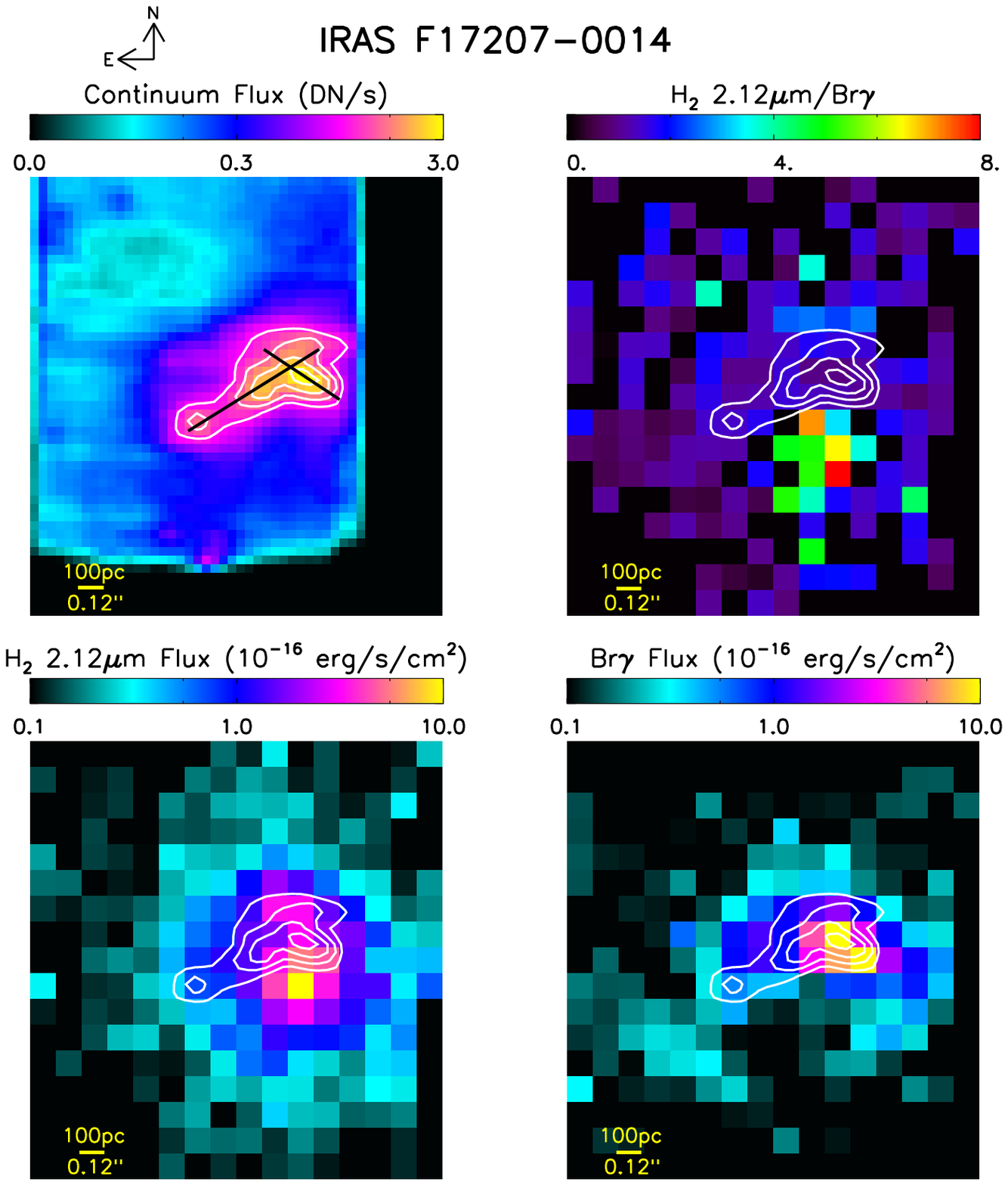}
\caption{Top Left: Continuum flux map of the central region of IRAS F17207-0014 at high spatial sampling (0\farcs035 spaxel$^{-1}$) in the $Hn4$ band \textbf{(1.652--1.737 $\mu$m)}. The major axes of the two nuclear disks identified in \citet{nucleardisks} are overlaid in black for context. Top Right: \molhy/\brg~flux ratio calculated from two bottom panels, with continuum contours overlaid; ratios indicate strong shocks ($>4$) to the south surrounded by a diffuse background of moderate shocks ($>0.6$). Bottom Left: \molhy~2.12 $\mu$m flux map (spatial sampling 0\farcs1 spaxel$^{-1}$) with contours from continuum overlaid. Bottom Left: Bottom Right: \brg~flux map (spatial sampling 0\farcs1 spaxel$^{-1}$) with continuum contours overlaid.}
\label{OSmaps}
\end{figure}

In Figure~\ref{OSmaps} we show the emission map of the 2.12 $\mu$m \molhy~line (bottom left panel) and \brg~line (bottom right panel), with contours at higher spatial sampling overlaid. We see that the \molhy~emission does not trace the \brg~or continuum emission (top left panel and white contours everywhere), and indeed that the \molhy/\brg~ratio (top right panel) confirms that the bulk of the \molhy~emission is likely due to shocks. The \molhy/\brg~ratio ranges from $\sim$4--8 below the intersection of the two disks; we additionally find that most of the field-of-view has an \molhy/\brg~ratio above 0.6, suggesting lower-level shocks could be widespread. The \molhy~velocity dispersion is slightly but not significantly elevated in the shocked region; across the field, most spaxels show velocity dispersion $>$100 km s$^{-1}$.

\subsubsection{Characteristics of the Shocked Region}

The region of highest \molhy/\brg~ratio ($\sim$4--8) extends at least 0\farcs4 (340 pc) from North to South and 170--340 pc from East to West. 

The total \molhy~and \brg~fluxes from this region are $9.75 \times 10^{-16}$ and $2.28 \times 10^{-16}$ erg s$^{-1}$ cm$^{-2}$ respectively, producing an average \molhy/\brg~ratio of 4.3. These fluxes are not corrected for extinction. The 1--0 S(1) \molhy~line flux we measure is approximately two orders of magnitude larger than the largest flux predicted by the shock models of \citet{ShullHollenbach78}. \citet{Chernoff82} found increased \molhy~emission produced by shock models including a magnetic field; our measured line intensity is about one order of magnitude larger than their prediction and covers approximately half of their simulated beam. These models were tuned to match the shocked molecular hydrogen emission surrounding the Orion Molecular Cloud; to our knowledge, a comprehensive grid of shock models does not exist. Therefore, the discrepancy in line intensity between our observations and these shock models may be due to differing shock conditions. The limited number (and, perhaps, limited applicability to this case) of shock models in the literature prevents us from drawing conclusions about the shock properties from this measurement at this time.

\begin{figure}
\centering
\includegraphics[trim= 1cm 0cm 0cm 0cm, scale=1.]{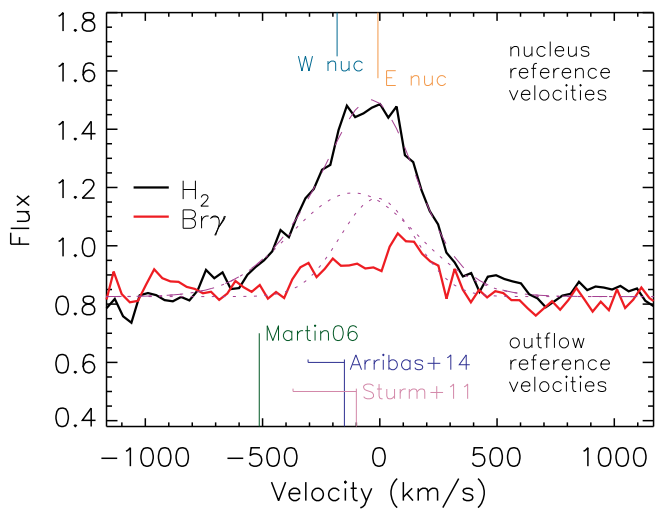}
\caption{\molhy~(black) and \brg~emission (red) as a function of velocity relative to systemic ($cz$=12,900 km s$^{-1}$).  The \molhy~emission is decomposed into two Gaussian fits (dotted purple lines: individual components, dashed purple line: total fit).  Reference velocities are shown for the two disks (from our kinematic modeling paper, Medling et al. submitted) along the top and for a selection of outflow velocities from the literature along the bottom: traced by Na ID absorption \citep{Martin06}, optical emission lines \citep{Arribas14}, and OH 79$\mu$m absorption \citep{Sturm11}.}
\label{h2vel}
\end{figure}

The velocity dispersion of the \molhy~emission, integrated over the shocked region, is 200 km s$^{-1}$; individual spaxels within the shocked region range from 145--200 km s$^{-1}$.  When integrating over the shocked region, the signal-to-noise ratio is sufficient to decompose the \molhy~profile into two Gaussian components (see Figure~\ref{h2vel}).  This produces a central Gaussian of $v_{offset} = -11\pm25$ km s${-1}$ and $\sigma = 159\pm56$ km s$^{-1}$ containing 36\% of the flux and a blueshifted Gaussian of $v_{offset} = -113\pm31$ km s${-1}$ and $\sigma = 268\pm24$ km s$^{-1}$ containing the remaining 64\% of the flux.

\section{Discussion}
\label{discussion}

Optical seeing-limited integral field spectroscopy from WiFeS (Section~\ref{wifes}), covering the central $\sim$20 $\times$ 30 kpc, reveals line ratios indicative of a mixing sequence of shocks. The data are consistent with mixing sequences of metallicities log(O/H)+12=8.99--9.17. In both sequences, the central spaxels show the lowest relative contribution to shocks; the fraction of emission line luminosity contributed by shocks in these cases is 10\% $\pm$ 10\% to 30\% $\pm$ 10\%, depending on metallicity. We also see evidence of two levels of shocks in the near-infrared AO-assisted integral field spectroscopy from OSIRIS (Section~\ref{osiris}), indicated by moderate and high \molhy/\brg~ratios (0.6--4 and 4--8, respectively).  It is encouraging that both sets of data show evidence of shocks in the central region, though direct comparison of shock properties is not currently possible due to the lack of near-infrared shock models. Nonetheless, it is clear that shocks penetrate both ionized and warm molecular gas in this system.

Although both datasets show conclusive evidence of shocks, it is challenging to derive the shock energetics. The WiFeS data presented here lack sufficient signal-to-noise to decompose into a shocked and non-shocked components, and we defer the reader to \citet{Soto12b} for estimates of the energy associated with the outflow traced by optical emission lines. In the most extreme shocks modeled, \citet{Draine83} estimates that up to 10\% of mechanical energy in a shock can be converted to \molhy~2.12$\mu$m emission. Using our \molhy~luminosity, this puts a lower limit on the mechanical energy input rate in the shock of $\sim$3$\times10^{39}$ erg s$^{-1}$. We note, however, that the efficiency of converting mechanical energy to \molhy~luminosity varies by more than five orders of magnitude in their models, based on shock velocity, magnetic fields, and pre-shock gas density and ionization fraction. Without a comprehensive grid of shock models predicting \molhy~luminosities and line ratios, we cannot constrain this further.

The complex nuclear dynamics of the central region, resolved by our OSIRIS spectra, makes it difficult to determine how the \molhy~emission relates kinematically to the nuclei. In Figure~\ref{h2vel} we show the \molhy~and \brg~emission lines summed over the shocked region as a function of velocity. We then overplot the nuclear velocities for the East and West disks, from our paper on kinematic modeling (Medling et al. submitted), as well as a selection of outflow velocities found in previous studies for reference. The range of outflow velocities found in the literature spans the \molhy~velocities. However, it is important to remember that the outflow velocities from the literature were found in very different spatial regions of the galaxy and/or at poorer spatial resolution; thus, matching velocity profiles is not sufficient for identifying a multiphase outflow. We also note that it is not precisely clear what velocity would be expected from shocked gas resulting from a nuclear collision: it could be shocked in place, maintaining a radial velocity between that of the two nuclei, or the collisional heating may drive the gas outwards, appearing blueshifted \citep[as in the models of][]{Cox04,Cox06}.

In the following sections we discuss three possible causes for the shocks seen in IRAS F17207-0014: AGN-driven outflows or heating, starburst-driven winds, and cloud-cloud collisions from the merging of the two progenitor galaxies.  We evaluate the evidence for each cause for each set of shocks below.

\subsection{Shocks Caused by AGN-Driven Outflows or X-Ray Heating}
\label{AGN}

AGN are capable of driving atomic and molecular gas out from the nuclear regions at moderate \citep[e.g. $\sim$350 km s$^{-1}$ in NGC~1266;][]{Alatalo11} to high velocities \citep[$\sim$2000 km s$^{-1}$;][]{Spoon13}. Such outflows are thought to be driven by radiation pressure from the AGN itself \citep{SilkRees98} and may be responsible for the existence of black hole scaling relations \citep*[e.g.][]{Springel05, DiMatteo05}. It is therefore plausible that shocks in the core of (and, perhaps, throughout) a ULIRG be caused by an AGN-driven wind. If an AGN were present, it's also possible that the resulting X-rays could heat the \molhy~gas, producing the high \molhy/\brg~ratios that we see.

In the case of IRAS F17207-0014 we find no evidence for an AGN from our data or from optical \citep{Yuan10}, mid-infrared \citep{Stierwalt13}, or X-ray \citep{Iwasawa11} observations. We cannot, however, rule out a Compton-thick AGN obscured to our line-of-sight but affecting gas visible along the plane of the sky. If such an AGN were driving an outflow causing these shocks we might expect to see high velocity gas as a result -- which is not present in either our data or in the literature \citep{Martin06,Soto12a, Soto12b, Rupke05a, Rupke05b}.  

Due to the lack of evidence supporting an AGN in this galaxy, we regard this scenario as unlikely.  

\subsection{Shocks Caused by Outflows Driven by Star Formation}
\label{winds}

Perhaps a more common driver of galactic winds is intense star formation activity \citep[see][for a review]{Heckman03}. The spaxels that are dominated by shocks in our WiFeS data appear to be in the outer regions of the galaxy (2--5 kpc), which is consistent with previous observations of galactic winds \citep{Rich10,Rich11,Rich14,Soto12a,Soto12b}. The line ratios in the central regions are plausibly low not because shocks are not present but because star formation dominates there \citep[e.g.][]{Armus89}.  

The signal-to-noise ratios of our spectra are generally too low to decompose the emission lines into a shocked and non-shocked component so it is not possible to say from these data whether the shocks we see in the WiFeS data are redshifted or blueshifted relative to the galaxy, or to measure a shock velocity. However, our results are consistent with those of \citet{Soto12a}, \citet{Soto12b}, and \citet{Arribas14} who also find line ratios increasing with galactic radius in a sample of ULIRGs and attribute them to outflows. The additional evidence for outflows in both Na I D absorption \citep{Martin06,Rupke05a,Rupke05b} and OH absorption \citep{Sturm11} makes it extremely likely that outflows are present on large scales.

The strong shocked region of elevated \molhy/\brg~flux seen below the two disks could be the origin of such an outflow. As this region points south, it is aligned with one of the WiFeS regions of increased shock fraction and consistent with the spectra of \citet{Soto12a}, which show the strongest shocks to the south; this could imply that we are seeing the base of an outflow that extends to $\sim5$ kiloparsecs. Indeed, the broader \molhy~component in Figure~\ref{h2vel} is blueshifted relative to the East disk with velocity offsets similar to the outflow velocities in the literature. Observations of other outflows beginning as such collimated structures show that they can remain mostly collimated over kiloparsec scales \citep[as in][]{Shopbell98}. Other regions of high shock fraction in the WiFeS data could be caused by a second outflow unseen in the OSIRIS data, perhaps due to geometry. We calculate the mechanical energy rate available in such an outflow as E$_{outflow}$ = $\dot{M} v_{out}^2 = 5.1\times 10^{40}$ erg s$^{-1}$, using $\dot{M}$ = 4 $M_{\sun}$ and $v_{out}$ = 200 km s$^{-1}$ from \citet{Soto12b}. This available mechanical energy rate exceeds the minimum mechanical energy rate ($3\times10^{39}$ erg s$^{-1}$, see Section \ref{discussion}) required to produce our \molhy~emission in shocks, depending on the shock parameters.

Most of the OSIRIS field of view shows \molhy/\brg~ratios of $\sim$2, still indicative of shocks, though less strong.  In the outflow scenario, one might imagine that uncollimated winds contribute to this moderate \molhy/\brg~level.

Some galaxies with outflowing winds show a correlation between optical emission line ratios and velocity dispersion \citep[e.g.][]{Rich10, Rich11,Ho14}, but we see no such trend here. It is possible that the lack of a correlation simply means that the shock velocity is not related to the turbulent velocity of the gas. It is worth noting that the velocity dispersions that we measure do not extend down to the few tens of km/s consistent with HII-region emission, possibly due to a combination of geometry and/or beam-smearing, which may affect our ability to detect such a correlation.  

\subsection{Shocks Caused by the Merger of Two Galaxies}
\label{merger}

The discovery of two nuclear disks in the core of IRAS F17207-0014 so close in projection \citep{nucleardisks} leads us to consider a third possibility, that some of the shocks seen are due to collisions of gas associated with the two disks as they are undergoing a close passage \citep[as in][]{Harwit87,Cox04}. The morphology of the strong ($>$4) \molhy/\brg~region in the OSIRIS data is consistent with the interface between the two disks.  

\citet{Rich11} found that the enhanced optical line ratios from shocks are easily washed out by star formation, and are thus easier to see on the outskirts where star formation rates are lower. It is important to note, however, that this does not mean that shocks are only present on the outskirts. The situation here is analogous: if the disk collision has caused the shocks, we still may not expect to see high \molhy/\brg~ratios directly on top of the two disks, where star formation is high and could wash out their signature. Seeing a stronger signal of shocks along the interface but below the increased star formation in the disks is consistent with this.

Given that IRAS F17207-0014 is in a late stage of merger \citep[merger class 5 out of 6;][]{Kim13}, it is plausible to imagine that cloud-cloud collisions are not uncommon, and may be widely distributed. These may also be responsible for the moderate levels of \molhy/\brg~($\sim$0.6--4) seen in the regions surrounding the nuclei in the OSIRIS data, independent of the cause of the highest \molhy/\brg~ratio. As the OSIRIS field-of-view is completely filled by the overlap region of the two progenitor galaxies, such collisions in the interstellar medium would appear as the diffuse background of shocks observed.

The collision of two disks of this size would produce a substantial amount of mechanical energy. We approximate the energy of the system as $T = \frac{1}{2} W = \frac{G M_{disk1} M_{disk2}}{r} = 4.5\times10^{58}$ erg, where G is the gravitational constant, $M_{disk1} = 9.8 \times 10^{9} M_{\sun}$, $M_{disk2} = 4.56 \times 10^{9} M_{\sun}$ from our dynamical modeling in Medling et al. (submitted), and the distance between the two disks r=200 pc from \citet{nucleardisks}; we divide this energy by 100 Myr as a typical timescale for nuclei to merge, in order to produce the average energy input from this source: $1.4 \times 10^{43}$ erg s$^{-1}$. This far exceeds the required amount of energy ($3\times10^{39}$ erg s$^{-1}$, see Section \ref{discussion}) to produce the measured \molhy~emission in shocks, depending on the shock parameters.

\subsection{Two Mechanisms Contributing to Shocks}
\label{twomech}

Both the starburst-driven wind scenario (Section~\ref{winds}) and the merger-driven collisional shock scenario (Section~\ref{merger}) can plausibly describe the data presented in this paper. We propose a hybrid scenario, in which the diffuse shocks (\molhy/\brg~ratio $\sim$0.6--4) filling most of our OSIRIS field-of-view are caused by the collision of the interstellar media associated with the two nuclei, and the strong shocks (\molhy/\brg~ratio $\sim$4--8) to the south of the nuclei trace the base of the outflow extending several kiloparsecs, visible in the wide-field optical spectroscopy. If the strong shocks do indeed show the base of a starburst-driven wind, then deeper optical integral field spectroscopy should find the strongest outflow signatures to the south.

This paper gives a first look at the unique perspective offered by combining a high-resolution view of a galaxy's core with the large-scale information of the system at large. This combination is an efficient way to look at the fueling and feedback of star formation and AGN activity.  We are in the early stages of a WiFeS+OSIRIS survey of nearby U/LIRGs, enabling us to trace spatially-resolved excitation mechanisms from small scales to large  and study the physical processes involved in star-forming and active galaxies and their effects on galaxy evolution.

\section{Summary}
\label{summary} 
We have presented additional analysis of integral-field spectroscopic observations of the ULIRG IRAS F17207-0014, a nearby late-stage major merger, making use of wide-field optical emission line ratios and complementary high spatial resolution AO-assisted near-infrared IFU data. We find:
\begin{itemize}
\item{The optical emission line ratios \nii/\ha, \oiii/\hb, \sii/\ha, and \oi/\ha~are elevated relative to pure star formation, and show a mixing sequence of star formation and shocks. These shocks are wide-spread and have their strongest relative contributions in the outskirts of the galaxy, which is suggestive of galactic-scale winds.}
\item{High spatial resolution near-IR imaging and kinematics reveal that the core of this system is a collision of two nuclear disks. High \molhy/\brg~ratios (4--8) extend from the interface between the disks to the south. More moderate \molhy/\brg~ratios (0.6--4) are present across most of the core, indicating the presence of shocks are widespread.} 
\item{We find plausible evidence for two potential causes of the shocks seen in IRAS F17207-0014. We propose that the diffuse low-level shocks are caused by cloud-cloud collisions driven by the merger, while the strong shocks towards the south trace the base of the starburst-driven wind seen in both our optical spectroscopy and that of other authors.}
\item{Pairing wide-field IFU data with high spatial resolution IFU observations of the central regions is a unique way to characterize the physical mechanisms present in the galaxy. We intend to collect this combination of data for a large sample of nearby (U)LIRGs in order to determine where and how galactic winds are launched, and how often they contribute to shocks across the galaxy.}
\end{itemize}

  
\section*{Acknowledgements}
We enthusiastically thank the staff of Siding Spring Observatory and the W. M. Keck Observatory and its AO team, for their dedication and hard work. The authors wish to pay their respects to the Elders - past, present and future - of the traditional lands on which Siding Spring Observatory stands, and to those of Hawai'ian ancestry on whose sacred mountain we are privileged to be guests. Without their generous hospitality, these observations would not have been possible. 
AMM, LJK, and MAD acknowledge the support of the Australian Research Council (ARC) through Discovery project DP130103925. This work was also partially funded by the Deanship of Scientific Research (DSR), King Abdulaziz University, under grant No. (5-130/1433 HiCi). The authors acknowledge this financial support from KAU. AMM and CEM also acknowledge support by the National Science Foundation under award number AST-0908796.


\bibliographystyle{mn2e}


\end{document}